%% file: hybrid-e.tex
\documentclass[fleqn,11pt]{article}
\usepackage{amsmath,amssymb,amsfonts,graphicx}

\input preamble.tex

\def\eqn#1{\eq\eqref{#1}}
\def\rf{\eqref}
\def\mn{_{\mu\nu}}
\def\MN{^{\mu\nu}}
\def\mN{_\mu^\nu}

\def\cR{{\cal R}}

\def\kappa{\varkappa}

\def\hG{{\hat\Gamma}}
\def\R{{\mathbb R}}
\def\oC{{\bar C}}
\def\og{{\bar g}}
\def\ophi{{\bar\phi}}
\def\oR{{\bar R}}

\def\ME{\mathbb{M}_{\rm E}}
\def\MJ{\mathbb{M}_{\rm J}}

\def\sph{spherically symmetric}
\def\ssph{static, spherically symmetric}

\def\bh{black hole}
\def\bhs{black holes}
\def\wh{wormhole}
\def\whs{wormholes}
\def\asflat{asymptotically flat} 

\usepackage{color}

\begin{document}
\thispagestyle{empty}
\twocolumn[

\bigskip

\Title {Spherically symmetric black holes and wormholes \yy
           in hybrid metric-Palatini gravity}
	
\Author{K. A. Bronnikov\foom 1}
	{\small
	Center fo Gravitation and Fundamental Metrology, VNIIMS, Ozyornaya ul. 46, Moscow 119361, Russia;\\
	 Institute of Gravitation and Cosmology, RUDN University, 
		ul. Miklukho-Maklaya 6, Moscow 117198, Russia;\\
	National Research Nuclear University ``MEPhI'', 
		Kashirskoe sh. 31, Moscow 115409, Russia}

\Abstract
   {The so-called hybrid metric-Palatini theory of gravity.(HMPG), proposed in 2012 by T. Harko et al., 
   is known to successfully describe both local (solar-system) and cosmological observations.  This 
   paper gives a complete description of \ssph\ vacuum solutions of HMPG in the simplest case where 
   its scalar-tensor representation has a zero scalar field potential $V(\phi)$, and both Riemannian ($R$) 
   and Palatini ($\cal R$) Ricci scalars are zero. Such a scalar-tensor theory coincides with general 
   relativity with a phantom conformally coupled scalar field as a source of gravity. Generic \asflat\ 
   solutions either contain naked central singularities or describe traversable \whs, and there is a 
   special two-parameter family of globally regular \bh\ solutions with extremal horizons. 
    In addition, there is a one-parameter family of solutions with an infinite number of extremal horizons 
    between static regions and a spherical radius monotonically changing from region to region. 
    It is argued that the obtained \bh\ and \wh\ solutions are unstable under monopole perturbations. 
    As a by-product, it is shown that a scalar-tensor theory with $V(\phi)=0$, in which there 
    is at least one nontrivial ($\phi \ne\const$) vacuum solution with $R\equiv 0$, necessarily reduces 
    to a theory with a conformal  scalar field (the latter may be usual or phantom).
    }

] 
\email 1 {kb20@yandex.ru} 

\section{Introduction}

  The century-old general relativity (GR) theory is still successfully passing all local observational 
  tests but faces well-known difficulties in describing large-scale phenomena, such as rotation
  curves of galaxies and clusters of galaxies and the cosmological evolution (these difficulties are
  frequently designated as the Dark Matter and Dark Energy problems). One of the ways of 
  addressing these problems is to assume the existence of still unobserved kinds of matter
  in the framework of GR, for example, weakly interacting massive particles (WIMPs) as dark 
  matter, a cosmological constant or a scalar ``quintessence'' field as dark energy, etc.
  \cite{DE}. Another way is to generalize or modify GR, invoking more general gravitational 
  Lagrangians than in GR (as, for instance, in $f(R)$ theories), additional degrees of freedom 
  (such as fundamental scalar fields in scalar-tensor theories), extra dimensions or new geometric
  quantities like torsion and nonmetricity, etc. \cite{ex-GR1, ex-GR2}.

  The recently proposed \cite{har12} hybrid metric-Palatini gravity (HMPG) theory belongs 
  to the second approach, sometimes called ``dark gravity'' to distinguish it from the ``dark energy''  
  approach. that remains in the framework of GR. In HMPG, one separately introduces the 
  Riemannian metric $g\mn$ and the independent connection $\hG\mn^\alpha$
  considering the total action \cite{har12}
\beq  				\label{S}
	S = \frac{1}{2\kappa^2}\int d^4 x\sqrt{-g} [R + F(\cR)] + S_m,
\eeq    
  where $R = R[g]$ is the Ricci scalar corresponding to $g\mn$, $F(\cR)$ is a function
  of the scalar $\cR = g\MN \cR\mn$, with the Ricci tensor $\cR\mn$ built in the standard way 
  from the connection $\hG\mn^\alpha$; as usual, $g = \det(g\mn)$, $ \kappa^2$ is the 
  gravitational constant, and $S_m$ is the action of nongravitational matter.
  
  Since the action \rf{S} is varied in both $g\MN$ and $\hG\mn^\alpha$, this theory 
  combines the metric and Palatini approaches to gravity and thus extends the $f(R)$
  theories. It has been shown that this theory does not contradict the local tests of gravity
  in the Solar system \cite{cap13b}, fairly well describes the dynamics of galaxies and 
  galaxy clusters thus addressing the dark matter problem \cite{cap13a}, and is able to 
  explain the modern cosmological observations that tell us about the properties of an 
  accelerating Universe. A detailed exposition of HMPG and its achievements may be 
  found in the reviews \cite{cap15,har18}. An even more general theory, containing an 
  arbitrary function of two variables, $R$ and $\cR$, is presented in \cite{boh13},
  see a recent further work on this theory in \cite{mont19}.
  
  This paper is devoted to a study of exact \ssph\ solutions of HMPG in the simplest case 
  where $F(\cR) \propto \cR$.This version of HMPG is equivalent to a scalar-tensor theory 
  (STT) of gravity \cite{har12,cap15} with zero scalar field potential $V(\phi)$, and 
  this theory coincides with GR with a source in the form of a phantom conformally invariant 
  scalar field. The solutions are obtained in the standard way, by reducing the STT to
  the Einstein conformal frame, in which they are well known (the so-called anti-Fisher 
  family of solutions). Many of the resulting HMPG solutions possess naked singularities, 
  and, due to a phantom nature of the effective scalar field $\phi$, there are generic \wh\ 
  solutions. In addition, under some special relation between the integration constants,
  there are nonsingular \bh\ solutions and a solution with infinitely many Killing horizons,
  each having its own radius. 
    
  The paper is organized as follows. The next section discusses the STT representation of 
  HMPG \cite{har12,cap15} and its basic features. It is shown, in particular, that if the STT
  with $V \equiv 0$ contains at least one nontrivial (that is, with $\phi\ne\const$) vacuum solution 
  where $R \equiv 0$, this theory describes a conformally invariant scalar field.  
  Sec. 3 describes \ssph\ solutions in this massless case ($V(\phi) =0, \ F(\cR) \propto \cR$)
  with special attention to globally regular solutions containing Killing horizons. These are 
  a two-parameter family of \asflat\ \bhs\ and a one-parameter family of geometries with
  an infinite number of horizons. There is also a discussion of possible \bh\  solutions 
  with nonzero potentials, and a brief consideration of the stability of the 
  presently obtained solutions. Sec. 4 contains some concluding remarks.  
  
\section {Some basic features of HMPG}  
  
  Varying the action separately in $g\MN$ and $\hG\mn^\alpha$, one obtains the field 
  equations according to which \cite{cap15, dan19} $\hG\mn^\alpha$ is a Riemannian 
  connection corresponding to the metric $h\mn = \phi g\mn$, conformal to $g\mn$,
  and, as a result, there is only one dynamic degree of freedom in addition to $g\mn$, 
  expressed in the scalar field $\phi = F_\cR \equiv df/d\cR$. The whole theory can then 
  be reformulated in terms of $g\mn$ and $\phi$ as a scalar-tensor theory with the 
  gravitational part of the action  \cite{har12,cap15}
\beq                           \label{S1} 
	S_g =  \int d^4x \sqrt{-g}\bigg[(1+\phi)R - \frac {3}{2\phi}(\d\phi)^2 - V(\phi)\bigg] 
\eeq          
  where\footnote
  	{Unlike \cite{har12, cap15, dan19} and many other papers, we are using here the metric 
  	 signature $(+  -   -\  -)$, therefore, the sign before $(\d\phi)^2 = g\MN \phi_{\mu}\phi_{\nu}$ 
  	 is different from that in the cited papers (the minus here corresponds to a phantom),
  	 while the meaning of \rf{S1} remains the same as there. We will also safely omit the factor 
  	 $1/(2\kappa^2)$ before the gravitational part of the action integral because we will only 
  	 deal with vacuum configurations where $S_m =0$.} 
    the scalar field potential is related to $f(\cR)$ by
\beq                             \label{VR}
		V(\phi) = \cR F_\cR - F(\cR).
\eeq    
  The action \rf{S1} corresponds to the original formulation of the HMPG in \cite{har12}.
  In the review paper \cite{cap15} the authors introduce one more constant $\Omega_A$
  into the action, so that in \rf{S1} one has $(\Omega_A + \phi)$ before $R$ instead of 
  $(1+\phi)$. They further remark that in the limit $\Omega_A \to 0$ the theory \rf{S} turns 
  into pure Palatini $F(R)$ gravity in which the scalar field has no dynamics (precisely as 
  in the Brans-Dicke theory with $\omega = -3/2$, to which \rf{S1} is converted if 
  one replaces $(1+\phi \mapsto \phi $), while in the limit $\Omega_A \to \infty$ the theory
  \rf{S} turns into metric $F(R)$ gravity. In what follows we adhere to the original STT 
  formulation \rf{S1}.

  It is easy to notice that \eqn{S1} represents a special case of the Bergmann-Wagoner-Nordtvedt 
  STT \cite{berg68, wag70, nor70} whose gravitational action reads
\beq   	 	\label{S-STT}
	S_g =\int\! d^4x \sqrt{-g}\Big[f(\phi)R  + h(\phi)(\d\phi)^2 - V(\phi)\Big],
\eeq     
  such that 
\beq  			\label{f,h-I}
	f(\phi) = 1+\phi, \cm     h(\phi) = - \frac{3}{2\phi}. 	
\eeq 
    
   In the general theory \rf{S-STT}, there is a standard transformation \cite{wag70}
   from \eqn{S-STT} (describing what is called the Jordan conformal frame of the STT) to the 
   Einstein conformal frame which is free from a nonminimal coupling 
   between the scalar field and the space-time curvature:
\bearr 			\label{J-E}
	\og\mn = f(\phi) g\mn, \qquad  \frac {d\phi}{d\ophi} = f (\phi) | D(\phi)|^{-1/2},
\nnn	
	D(\phi) = f(\phi)h(\phi) + \frac 32 \bigg(\frac{df}{d\phi}\bigg)^2,
\ear
  resulting in a simpler form of the action:
\beq  	\nhq		\label{S-E}
	S_g = \int \! d^4x \sqrt{-\og}
		\bigg[\oR + n \og\MN \ophi_{,\mu}\ophi_{,\nu} - \frac{V(\phi)}{f^2(\phi)}\bigg],
\eeq                      
  where bars mark quantities obtained using the transformed metric $\og\mn$,
  and $n = \sign D(\phi)$. In the theory \rf{S} under consideration, we have
\beq                         \label{tan}
	  D = -\frac {3}{2\phi}, \quad\ n = -1, \quad\ \phi = \tan^2 \frac{\ophi}{\sqrt 6}  
\eeq      
  We have to put $n=-1$ since $\phi > 0$ by construction. It is a very important point,
  indicating a phantom nature of the scalar field $\ophi$ and hence $\phi$ which, 
  in particular, favors the existence of wormholes. 
  
  The transition \rf{J-E} from \rf{S-STT} to \rf{S-E} is a map from the Jordan-frame manifold
  $\MJ$ with the metric $g\mn$ to the Einstein-frame manifold $\ME$ with the metric 
  $\og\mn$. This map is completely reversible if the conformal factor $f(\phi)$ is regular 
  and turns neither to zero nor to infinity in the whole range of $\phi$ relevant to the problem 
  under study; in this case, one may assert that there is a single manifold equipped with
  two different Riemannian metrics $g\mn$ and $\og\mn$. If $f(\phi)$ is zero or infinity 
  at some value of $\phi$, it may happen that a singularity in $\ME$ maps to a regular
  surface in $\MJ$ (or vice versa), and $\MJ$ then has a continuation beyond this surface.
  This phenomenon, called {\it conformal continuation} \cite{kb02}, occurs in many known
  scalar-vacuum and scalar-electrovacuum solutions including black holes and wormholes
  with conformally coupled scalar fields \cite{kb70, kb73} as well as the so-called cold 
  black holes in Brans-Dicke theory \cite{CBH1, CBH2}.  
   
  Meanwhile, the transition \rf{J-E} is also a well-known efficient method of finding 
  exact or approximate solutions to the equations of the theory \rf{S-STT} due to the
  comparative simplicity of the action \rf{S-E}. I can hardly agree with the statement from
  \cite{dan19} ``A crucial mathematical requirement for transformations (28) [(6) in this paper] 
  to be valid is that they must be nonsingular for the considered range of the scalar field variable.''
  In fact, the transformation \rf{J-E} is, from the viewpoint of solving differential equations,
  just a substitution which can in principle work in only a part of the relevant range of 
  the variables involved. It can then happen that the solutions obtained with such a 
  substitution will be incomplete, covering only a part of the range of independent 
  variables or parameters, and the full range may be then obtained using analytical 
  extensions. Conformal continuations are just an example of such extensions. But in 
  any case a solution obtained using this substitution is a valid solution of the original 
  set of equations. Therefore the above-mentioned regularity is not a ``crucial 
  mathematical requirement'': its possible violation only requires due attention and study.   
  
  Even more than that: as is clear from \cite{kb73,CBH2}, conformal continuations emerge 
  only in special cases, for example, black holes and wormholes with a conformally 
  coupled scalar field \cite{kb73,visser00} are found at special values of the integration 
  constants, whereas in more general cases singularities in $\ME$  are mapped into
  singularities in $\MJ$ (maybe of another nature).

  As to the action in the form \rf{S1}, we notice that, as follows from \eqn{tan}, the whole 
  range of the $\phi$ field, $\phi \in \R_+$, is covered by a single suitable segment 
  of $\ophi$, for example, $\ophi \in (0, \sqrt{6}\pi/2)$, and the conformal factor 
  $f(\phi) = 1 + \phi = 1/\cos^2(\ophi/\sqrt{6})$ is regular in the same range. Therefore, 
  solutions to the field equations must be equivalently described in terms of $\phi$ 
  and $\ophi$, up to possible special cases with conformal continuations. We shall 
  clearly see that in the next section.

  One more important observation is in order: after the scalar field reparametrization in \rf{S1}, 
\beq
		\phi \mapsto  \chi, \qquad   \phi = \chi^2/6,
\eeq    
  the gravitational field action reads
\bearr                                     \label{S2}                          
		S_g = \int \! d^4x \sqrt{-g} \Big[(1+\chi^2/6)R 
\nnn \inch	\cm	
		- (\d\chi)^2 - W(\chi)\Big],
\ear
  which is nothing else but the action of GR with a source of gravity in the form of a 
  conformally coupled scalar field (as mentioned in \cite{cap15}) of phantom nature,
  with the potential $W(\chi) = V(\phi)$. Curiously, this version of STT was as early as in 1970
  considered by Zaitsev and Kolesnikov  \cite{ZK} as a viable alternative to GR in cosmological 
  and astrophysical applications. 
   
  In the case of a massless field $\phi$, $V(\phi) =0$, it follows from the field equations due 
  to \rf{S1} or \rf{S2} that any vacuum solution ($S_m=0$) has a zero Ricci scalar, $R =0$ 
  (see \cite{dan19}). However, one can prove an inverse general result:
 
\medskip 
   {\it If there is a vacuum solution with $R \equiv 0$ and a non-constant scalar field in a theory 
   \rf{S-STT}, with $V \equiv 0$, then this STT reduces either to GR with a conformally 
  coupled scalar field or to pure conformal field theory.} 

\medskip    
  Indeed, consider a general STT with $V(\phi) =0$ in the parametrization of the kind \rf{S2}, 
  that is,  
\beq                      \label{S3}
		S_g = \int   d^4x \sqrt{-g} \Big[ f(\phi) R + \eps (\d\phi)^2\Big],      
\eeq
  where $\eps = \pm 1$, so that $\eps =+1$ corresponds to a canonical scalar and 
  $\eps = -1$ to a phantom one. The vacuum field equations read
\bearr                          \label{e-S3}
	-\Half \delta\mN f R + (R\mN + \nabla_\mu \nabla^\nu - \delta\mN \Box)f 
\nnn \cm \cm
		+ \eps \phi_{,\mu} \phi^{,\nu} - \Half \eps \delta\mN (\d\phi)^2 =0,
\yyy		                    \label{ef-S3}
	2\ \Box\ \phi = \eps f_\phi R, 	
\ear			    
   where $\Box = g\MN \nabla_\mu \nabla_\nu$, and $f_\phi = df/d\phi$ The trace of \rf{e-S3} is
\beq                            
		\eps (\d\phi)^2  + fR + 3\ \Box\, f =0,
\eeq              
  Expressing $\Box\, f$ in terms of the derivatives of $\phi$ and using \rf{ef-S3} for 
  $\Box\,\phi$, we obtain
\beq
		(\eps + 3 f_{\phi\phi})(\d\phi)^2 + \Big(f + \frac 32 \eps f_\phi^2 \Big) R =0.
\eeq                  
  Now, assuming a nontrivial $\phi(x)$, such that $(\d\phi)^2 \ne0$, let us look under which 
  conditions this equation is compatible with $R=0$: evidently, we should require
\beq                         \label{R__0}
  		\eps + 3 f_{\phi\phi} =0, 
\eeq  
  so that
\beq
		f_\phi = -\frac 13 \eps (\phi+C_1), \quad f = - \frac \eps 6 (\phi+C_1)^2 + C_2,
\eeq      
   with $C_{1,2} = \const$, or, denoting $\tilde\phi = \phi + C_1$, 
\beq
               f(\phi) = C_2 - \frac 16 \eps {\tilde\phi}{}^2, 
\eeq          
   which means that $f(\phi)$ corresponds to conformal coupling of the scalar $\phi$ to the metric,
   while $\phi$ is allowed to be canonical or  phantom. If $C_2 =0$, we are dealing with pure
   conformal field theory, and $C_2 \ne 0$ describes the presence of the Einstein-Hilbert term 
   in the action.
                                               
\section{Static, \sph\\ solutions}  

   Consider \ssph\ vacuum configurations in the theory \rf{S1}. Evidently, instead of solving
   the Jordan-frame field equations that directly follow from variation of \rf{S1} in $g\MN$ and 
   $\phi$, that is,
\bearr                          \label{e-J}
	(G\mN + \nabla_\mu \nabla^\nu - \delta\mN \Box)(1+\phi) 
\nnn \qquad
		- \frac 3{2\phi} \Big(\phi_{,\mu} \phi^{,\nu} + \Half \delta\mN (\d\phi)^2\Big) 
					+ \Half\, V =0,
\yyy		                    \label{ef-J}
	\frac 3\phi \,\Box \, \phi - \frac {3}{3\phi^2} (\d\phi)^2 + R - \frac {dV}{d\phi} =0,
\ear    
  ($G\mN = R\mN - \half \delta\mN R$ is the Einstein tensor),
  it is easier to solve the Einstein-frame equations that follow from \rf{S-E},
\bearr                       \label{e-E}
       	{\bar G}\mN - \ophi_{,\mu} \ophi^{,\nu} + \Half \delta\mN (\d\ophi)^2 + \Half \delta\mN W =0,
\yyy       		         \label{ef-E}  
      		2\,{\bar \Box}\, \ophi - \frac{dW}{d\ophi}=0, \qquad W(\ophi) := \frac {V(\phi)}{(1+\phi)^2}, 		                    
\ear
  and transform the solutions back to $g\mn$ and $\phi$ according to \rf{J-E}. As is stressed in
  \cite{dan19}, the HMPG theory is naturally formulated in terms of the Jordan-frame metric 
  $g\mn$ and the scalar field $\phi$.  
  
\subsection{Solutions for $V(\phi) =0$}

  In the case of a massless field ($V(\phi) \equiv 0$), the Einstein-frame solution is well 
  known: it is the phantom counterpart \cite{kb73,b-lei57,h_ell73} of Fisher's original 
  solution with a canonical massless minimally coupled scalar field \cite{fish48}. This 
  solution with a phantom scalar (often called the ``anti-Fisher'' solution) splits into three 
  branches, which can be presented in a unified way using the notations of \cite{kb73}. 
  Namely, consider the general \ssph\ metric in $\ME$
\beq                                \label{ds_E}
	  ds^2_{\rm E} = \e^{2\gamma}dt^2 - \e^{2\alpha}du^2 - \e^{2\beta} d\Omega^2,
\eeq    
  where $\alpha, \beta, \gamma$ are functions of an arbitrarily chosen radial 
  coordinate $u$, and  $d\Omega^2 = d\theta^2 + \sin^2 \theta\, d \varphi^2$ is the line 
  element on a unit sphere. Then, using the harmonic coordinate condition 
\beq
		\alpha(u) = 2\beta(u) + \gamma(u),
\eeq    
  one can write the anti-Fisher solution as follows \cite{kb73}:
\bearr                  \label{sol-E}
           \ophi = \oC u + \ophi_0,\qquad \gamma(u) = - hu, 
\nnn           
           e^{-\beta(u)-\gamma(u)} = s(k,u) := \vars {\!
                        k^{-1}\sinh ku,  \ & k > 0 \\
                                    u,  \ & k = 0 \\
                        k^{-1}\sin ku,   \ & k < 0,  }           
\nnn
	   h, \ k,\ \oC,\ \ophi_0 = \const,
\ear           
  where $u > 0$ while the integration constants $h$, $\oC$ (the scalar charge) and $k$ are 
  related by the equality
\beq                                                       \label{int-0}
            2k^2\sign k = 2h^2 - \oC^2.
\eeq
  Detailed descriptions of its properties may be found, e.g., in \cite{bbook12, kb-stab11}.\footnote
  		{The (anti-)Fisher metric is given by the expression in curly brackets in \rf{ds_J} and
  		corresponds to a minimally coupled scalar $\ophi$ or $\psi$ . The only difference 
  		between the solutions for a canonical scalar (Fisher's) and that for a phantom scalar 
  		(discussed here) is that in the canonical case there is a plus instead of a minus before 
  		$C^2$ in \rf{int-0} and \rf{int-1}; it leads to 	$k >0$. and so there is only branch A with
  		$a <1$ according to \eq (31) below. In the phantom case, $k$ has any sign, 
  		which leads to three branches.}
  
  Accordingly, the Jordan-frame metric and the scalar field $\phi$ in the theory \rf{S1}
  with $V(\phi)=0$ may be presented as
\bearr                   \label{ds_J}
             ds_J^2 = \cos^2 (Cu + \psi_0) \bigg\{ \e^{-2hu}dt^2 
\nnn \cm     \cm       
             - \frac{e^{2hu}}{s^2(k,u)}\bigg[\frac {du^2}{s^2(k,u)} + d\Omega^2 \bigg]\bigg\},
\yyy                       \label{phi}
  	     \phi (u) = \tan^2 \psi,  \qquad   \psi := \ophi/\sqrt{6} = Cu + \psi_0,     
\ear      
  where the notation  $\psi = \ophi/\sqrt{6}$ has been introduced for convenience,
  $C = \oC/\sqrt{6}$, and $\psi_0 = \ophi_0/\sqrt{6}$. The integration constants $k, h$ and
  $C$ are related by 
\beq 			\label{int-1}
		k^2 \sign k = h^2 - 3 C^2.
\eeq      
  Without loss of generality we assume $|\psi_0| < \pi/2$.
  
  A direct inspection shows that these $g\mn$ and $\phi$ do satisfy the field equations 
  \rf{e-J} and \rf{ef-J} with $V(\phi) \equiv 0$. It is also directly confirmed that the scalar 
  curvature is zero for $g\mn$ given by \rf{ds_J}, as should be the case for $V \equiv 0$.
  (According to \rf{VR}, $V = 0$ implies $\cR = \const\cdot R$, therefore in this case 
  $\cR$ is also zero \cite{dan19}.)
     
  Let us briefly describe the properties of this solution. To begin with, in all cases the
  metric \rf{ds_J} is \asflat\ at $u=0$ which corresponds to the spherical radius 
  $r \equiv \sqrt{-g_{\theta\theta}}$ tending to infinity, so that $r \propto 1/u$ as 
  $u\to 0$.\footnote 
  		{The conformal factor $\cos^2 \psi$ is not normalized to unity at $u=0$
  		if $\psi_0 \ne 0$, which, however,	 does not affect the further description.}
  A comparison with the Schwarzschild metric at small $u$ using a transition to the 
  coordinate $r$ leads to the following expression for the Schwarzschild mass:
\beq  			\label{m}  
  		m = h \cos \psi_0 + C \sin \psi_0.
\eeq  
  Other properties of the solution are different for different values of $k$, comprising 
  three branches according to the definition of $s(k,u)$ in \rf{sol-E}.
  
\medskip\noi
  {\bf Branch A:} $k > 0$. It is helpful to put 
\bearr                          \label{u-P}
		\e^{-2ku} = P(x) := 1 - \frac{2k}{x},  
\nnn   
		\e^{-2hu} = P(x)^a, \qquad     a = \frac hk = \pm \sqrt{1 +  \frac{3C^2}{k^2}}. 
\ear   
   The metric acquires the form 
\bearr                  \label{ds-A}
                ds_J^2 = \cos^2 \psi \Big[ P^a dt^2 - P^{-a} dx^2 - x^2 P^{-a} d\Omega^2 \Big],           
\nnn \cm  
		\psi = \psi_0 - \frac{C}{2k} \ln P. 
\ear      
   With the new coordinate $x$, flat spatial infinity corresponds to $x\to \infty$. As $x$ 
   decreases from infinity, $P(x)$ decreases beginning from unity, ultimately reaching 
   the value where $\cos\psi =0$ and, according to \eqn{tan}, $\phi \to \infty$. This happens  
   where $\ln P(x) = -(2k/C)(\pi/2-\psi_0)$ if $C > 0$ and where 
   $\ln P(x) = (2k/C)(\pi/2 + \psi_0)$ if $C < 0$. It evidently corresponds to finite $P$, that is, 
   $x > 2k$: it is a naked central (i.e., where the radius $r =0$)  singularity which is 
   attractive for test particles since $g_{tt} =0$.   
   
\medskip\noi
  {\bf Branch B:} $k = 0$. In this case, it is helpful to substitute $u = 1/x$, so that $x=\infty$
    corresponds to flat spatial infinity. The solution has the form
\bearr                     \label{ds-B}
	  ds_J^2 = \cos^2\psi \Big[ \e^{-2h/x}dt^2 - \e^{2h/x}(dx^2 + x^2 d\Omega^2) \Big],
\nnn	  
     		\psi = \psi_0 + C/x, \cm h^2 = 3C^2.
\ear               
   The coordinate $x$ ranges from $x_s$ to infinity, where $x_s$ is the value of $x$ 
   where $\cos\psi =0$ and $\phi = \infty$. We have $x_s = C/(\pi/2-\psi_0)$ if $C >0$  
   and $x_s = - C/(\pi/2 + \psi_0)$ if $C < 0$. In both cases, as in branch A, it is a central
   attractive singularity. 

\medskip\noi
  {\bf Branch C:} $k < 0$. In this case, it makes sense to use the original coordinate $u$
  (which is harmonic in the Einstein frame), and the solution reads
\bearr              \nhq              \label{ds-C}
	ds_J^2 =  \cos^2\psi \bigg[\e^{-2hu}dt^2 - \frac{k^2 \e^{2hu}}{\sin^2 ku}
\nnn \inch  \cm    \times	
	                        \bigg(\frac{k^2 du^2}{\sin^2 ku} + d\Omega^2\bigg) \bigg],
\nnn \qquad
                      \psi = \psi_0 + Cu. \cm h^2 = 3C^2 - k^2.				
\ear    
   As already mentioned, $u =0$ is flat spatial infinity with the Schwarzschild mass \rf{m}. 
   However, on the whole, the nature of the solution crucially depends on an interplay 
   between the constants $k$, $C$ and $\psi_0$, depending on which of the functions 
   $\sin |k|u$ or $\cos\psi$ is the first to vanish at growing $u$ beginning from zero.  
   Adhering to \asflat\ solutions, we require that at $u=0$ the conformal factor $\cos^2\psi$
   should be nonzero, therefore without loss of generality we require $|\psi_0| < \pi/2$. 
   
   Three possible behaviors should be singled out (we assume, for certainty, $C > 0$),
   see Fig.\,1.
   
\begin{figure}   
\centering
\includegraphics[width=6.5cm]{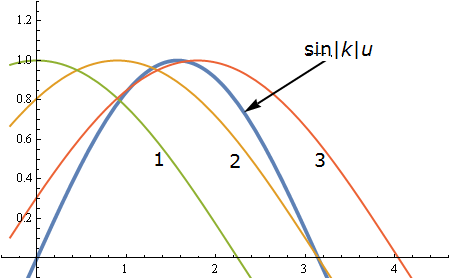}  
\caption{\small
	Origin of the behaviors C1, C2, C3 of the metric \rf{ds-C}. 
	We take for certainty $|k| = 1$; curves 1--3 plot $\cos\psi$ for $C = 0.7$ and 
	different $\psi_0$. The solution properties depend on which of the functions 
	$\sin |k|u$ or $\cos\psi$ is the first to reach zero at $u>0$. So curve 1 corresponds 
	to space-time with a naked singularity, curve 2 to a black hole, and curve 3 to a wormhole.  }       
\end{figure}   
\medskip\noi
  {\bf C1:} $(\pi/2 - \psi_0)/C < \pi/|k|$. The coordinate $u$ ranges from zero to 
  $u_s = (\pi/2 - \psi_0)/C$, at which $\cos\psi =0$, and $u=u_s$ is a central naked 
  singularity quite similar to the one in branches A and B. 

\medskip\noi
  {\bf C2:} $(\pi/2 - \psi_0)/C > \pi/|k|$. The coordinate $u$ ranges from zero to $u_*=\pi/|k|$
  where $\sin ku =0$, and there  occurs the second spatial infinity: as $u \to \pi/|k|$,  
  the radius $r \to \infty$ while $g_{tt}$ and $\phi$ remain finite.  This 
  second infinity is also flat, and the Schwarzschild mass\footnote
  		{As can be directly verified, for a general \asflat\ \ssph\ metric in the form \rf{ds_E},
  		with an arbitrarily chosen radial coordinate $u$, the Schwarzschild mass 
  		at a value $u=u_*$ corresponding to flat infinity is determined as
  	\[ 
  			m_* = - \lim\limits_{u \to u_*} (\e^\beta \gamma'/\beta'), 
  	\]
  		where the prime stands for $d/du$.}	   	   
  is there equal to 
\beq
                   m_* = -\e^{hu_*}(h \cos \psi_* + C \sin \psi_*),
\eeq    
  where $\psi_* = \psi_0 + C\pi/|k| < \pi/2$ is the value of $\psi$ at $u=u_*$.  So, it is a 
  \wh\  configuration, only quantitatively different from its anti-Fisher and Brans-Dicke 
  counterparts, see, e.g., \cite{kb73,brss10}. 
   
\medskip\noi   
  {\bf C3:}  $(\pi/2 - \psi_0)/C = \pi/|k|$. In this special case, at $u= u_1= \pi/|k|$ both
  $\sin |k|u$ and $\cos\psi$ turn to zero, while the spherical radius 
  $r = \sqrt{-g_{\theta\theta}}$ is finite but $\phi = \infty$. Near $u = u_1$, the metric 
  looks like
\bearr   \nq\,
		ds_J^2 = C^2 \bigg[\! \e^{-2hu_1} \Delta u^2 dt^2
			- \! \e^{2h u_1}\frac{du^2}{\Delta u^2} 
			-  \! \e^{2h u_1} d\Omega^2\bigg],
\nnn
\ear      
  where $\Delta u = u_1 - u$. It shows that $u=u_1$ is a double (extremal) horizon,
  and it is the only special case of the solution under study in which it describes a \bh.  
  
  Quite similar three cases are observed if $C < 0$, hence $\psi$ decreases at 
  increasing $u$, and $\cos\psi =0$ corresponds to $\psi = -\pi/2$.
  
 \subsection{Black holes with $V=0$} 
 
  The \bh\ solutions deserve a more detailed discussion.  
     
  The condition $(\pi/2 - \psi_0)/C = \pi/|k|$ leads to 	$\psi_0 = \pi (1/2 - C/|k|)$.
  The requirement $\psi_0 > - \pi/2$ then implies $C < |k|$.
  This inequality, obtained formally, has an evident meaning: since $\cos\psi \ne 0$ at 
  $u=0$, the plot of $\cos \psi$ must be wider than that of $\sin ku$ in order that their 
  first positive zeros coincide ($u=u_1$). It is then clear that, as the coordinate $u$ further 
  increases (that is, when we are looking what happens beyond the horizon), the next zero of 
  $\sin ku$, namely, $u_2 = 2\pi/|k|$ is reached earlier than a zero of $\cos\psi$. 
  
  This in turn completely reveals the global structure of our \bh\ configuration. Indeed, the 
  value $u = u_2$ with $\sin |k|u =0$ corresponds to a second spatial infinity, quite similar 
  to that in a \wh\ solution. Like the \whs\ described above, this space-time is globally regular,
  but now it is not two-side traversable because of the horizon. Its Carter-Penrose diagram 
  is shown in Fig.\,2.
  
\begin{figure}   
\centering
\includegraphics[width=5cm]{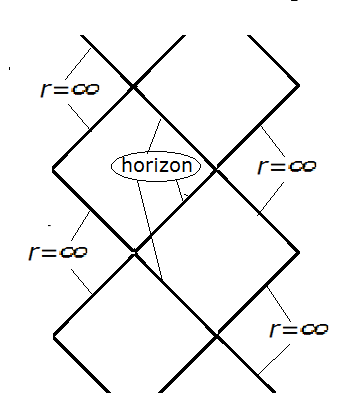}   :
\caption{\small
	Carter-Penrose diagram for regular \bh\ space-times with the metric \rf{ds-C}  
	The diagram infinitely extends up and down.     }
\end{figure}   
    
  It should be noted that the present  \bh\ solution has much in common with 
  the well-known solution for a \bh\ supported by a conformally coupled scalar 
  field \cite{kb70, kb73, bek74} with the metric
\beq                  \label{old_BH}
		ds^2 = \Big(1 - \frac mr\Big)^2 dt^2
					- \Big(1 - \frac mr\Big)^{-2}dr^2 - r^2 d\Omega^2.
\eeq    
  and $\phi = \sqrt{6} m/(r-m)$, $m= \const$ being the mass.  In both cases,
  
\medskip\noi   
  (i)  the \bhs\ are described by special solutions to the Einstein-scalar equations; 
  
\medskip\noi     
  (ii) all of them are \asflat\ and extremal and therefore have zero Hawking temperature; 
  
\medskip\noi     
  (iii) the supporting scalar fields are infinite on the horizon, but the effective 
  stress-energy tensor components $T\mN$ are finite there (as follows from finite values 
  of the Einstein tensor components $G\mN$); 
  
\medskip\noi     
  (iv) lastly, in both cases the scalar curvature is zero in the whole space. 
  
\medskip  
  However, there are important differences:
  
\medskip\noi  
  (i)  the solution \rf{old_BH} corresponds to a canonical scalar field in the Einstein frame 
  while for \rf{ds-C} such a field is phantom; 
  
\medskip\noi  
  (ii)  the solution \rf{old_BH} has a singular center $r=0$ (it has the same geometry as the 
  extreme Reissner-Nordstr\"om space-time) while the present \bh\ space-time has no 
  center and is globally regular;
  
\medskip\noi  
  (iii)  the solution \rf{old_BH} has only one free parameter $m$ while the present one 
  contains two independent parameters, for example, $k$ and $C$;   
  
\medskip\noi  
  (iv) the static region of the solution \rf{old_BH} can be obtained from Fisher's solution 
  \cite{fish48} in the Einstein frame only with the help of a conformal continuation \cite{kb73, kb02}, 
  whereas for the \bh\ case of \rf{ds-C} the Einstein-frame image of the horizon is the 
  second spatial  infinity of an anti-Fisher \wh. A conformal continuation is only required 
  for the extension beyond the horizon.
  
\medskip  
  One can notice that none of the \ssph\ solutions of the theory \rf{S1} with $V \equiv 0$ 
  possess simple horizons with finite temperature, contrary to the results announced in 
  \cite{dan19}.
            
\subsection{A geometry with infinitely many horizons}           

  A one-parameter family of geometries of interest is obtained if we abandon the 
  asymptotic flatness requirement and put 
\beq
		C = |k|, \qquad \psi_0 = -\pi/2,
\eeq    
  so that $\cos^2 \psi = \sin^2 ku$, and the Jordan-frame metric then reads
\beq  \nhq                     \label{inf}
		ds_J^2 = \sin^2 ku \e^{-2hu} dt^2 - k^2 \e^{2hu}
				\bigg( \frac{k^2 du^2}{\sin^2 ku} + d\Omega^2 \bigg),
\eeq    
  where $h = \pm \sqrt{2k^2}$ according to \rf{ds-C}. This space-time, with $u \in \R$,
  is a union of an infinite number of static regions, each being described by a single half-wave 
  of the function $\sin ku$, and double horizons between them at each $u = \pi n/|k|$, with 
  any integer $n$. A symbolic picture of this geometry is presented in Fig.\,3. Though, 
  the space-time $\MJ$ with the metric \rf{inf} is,  in a clear sense, ``much larger'' than 
  depicted since the spatial distance from any regular point in any static region to each of its 
  two nearest horizons is infinite due to divergence of the integral $\int \! \sqrt{|g_{uu}|} du$ 
  (which is a common feature of all double horizons). 
  
\begin{figure}   
\centering
\includegraphics[width=7cm]{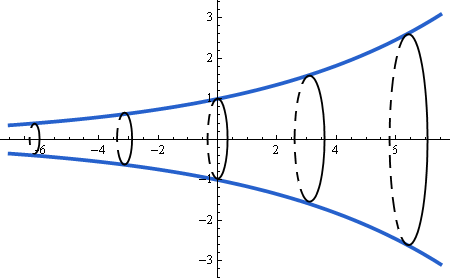}  
\caption{\small
	         A symbolic representation of the geometry \rf{inf} as a horn with cross-sections 
	         whose radius grows with growing $u$. For certainty, we have put $|k|=1$. 
	         The selected sections at $u = n\pi$ represent the horizon spheres.  
	}
\end{figure}   
  
  Thus the Jordan-frame manifold $\MJ$ maps to a countable number of
  Einstein-frame manifolds $\ME$, each of the latter being an anti-Fisher \wh\ whose
  both flat infinities map into horizons in $\MJ$. One more example of a construction 
  with an infinite number of conformal continuations was built in \cite{kb02}: there, it is 
  presented by a solution for a conformally coupled scalar field $\phi$ with the normal
  sign of kinetic energy and a nonzero potential $U(\phi)$.
  In that example, the continuation took place through ordinary surfaces $S_{\rm trans}$
  of finite radius, and the whole $\MJ$ had no horizons and was either completely static 
  or completely cosmological. In the first case, its shape was that of an infinitely long tube 
  with a periodically changing diameter. In the second case, $\MJ$ represented a 
  (2+1)-dimensional cosmology with a periodically (and isotropically) changing scale
  factor. Unlike that, in our case, all transition surfaces $S_{\rm trans}$ are double horizons
  between static regions, and the spherical radius monotonically changes from one region 
  to another. 
  
\begin{figure}   
\centering
\includegraphics[width=7cm]{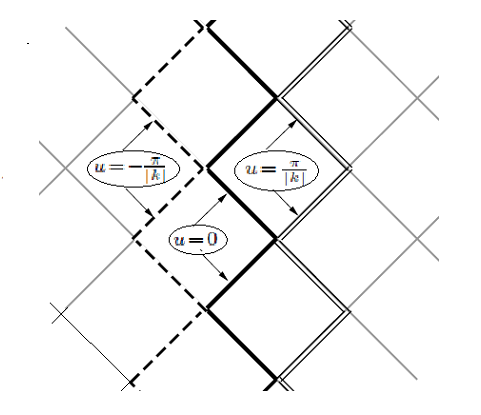}  
\caption{\small
	         Carter-Penrose diagram of the manifold $\MJ$ with the metric \rf{inf}. The diagram
	         occupies the whole plane. Horizons with $u = -\pi/|k|,\ 0,\ \pi/|k|$ are shown by the 
	         dashed, thick and double broken lines, respectively. 
	}
\end{figure}   

  The global causal structure of the manifold $\MJ$ with the metric \rf{inf} is described by 
  the Carter-Penrose diagram that occupies the whole 2-plane and consists of equal squares
  (Fig.\,4). The borders between adjacent squares show the horizons. One can recall that 
  such diagrams in cases with multiple horizons frequently involve branching and contain 
  a lot of sheets: for example, a diagram with four simple horizons can be drawn on a single 
  plane but a configuration with one single and one double horizons requires a multi-sheet 
  diagram with overlapping regions \cite{we12}. However, branching takes place if some
  locations on the diagrams depict singularities or infinities, while in the present case 
  there is no branching since any point of the diagram in Fig.\,4 (including the knots) 
  corresponds to regular spheres, and there is no spatial infinity.
        
  Another example of a manifold with an infinite number of horizons was obtained among 
  solutions for phantom dilaton-Einstein-Maxwell \bhs\ in \cite{cle09}.
  
 \subsection{On solutions with nonzero potentials $V(\phi)$}     
 
  If $V(\phi) \ne 0$, it is not so easy to solve the field equations, and there is a modest number 
  of examples of \ssph\ solutions with some special potentials even in the case of minimally
  coupled scalar fields. For the latter, there exist some general theorems allowing for judging 
  on possible solution behaviors without having these solutions. These are, above all, the
  no-hair theorems (see, e.g, \cite{herd15} for a recent review) indicating the conditions that 
  exclude horizons, and the global structure theorems \cite{kb01} 
  which work when the no-hair theorems can be violated and tell us about the maximum possible 
  number of horizons: this number is two, and it is only one for \asflat\ configurations 
  
  What can be said on Jordan-frame metrics, which are conformal to Einstein-frame ones 
  with minimally coupled scalars? If the conformal factor $f(\phi)$ in the mapping \rf{J-E} is regular 
  and finite in its whole range, most of the theorems actually extend to $\MJ$ since (while mapping 
  to either side) a flat infinity maps to a flat infinity, a horizon maps to a horizon, and the scalar field 
  potential preserves its sign. However, if $f(\phi)$ is somewhere singular, there emerges the possibility
  of conformal continuations, and how mighty they can be, is demonstrated by the above example 
  of an infinite number of horizons in $\MJ$. An important restriction is that horizons coinciding 
  with transition surfaces at continuations are double (extreme): however, one cannot exclude that
  $\ME$ contains a simple horizon, but as a result of a continuation the new region of $\MJ$ maps
  to the same $\ME$, and it will then contain two simple horizons. We can conclude that $\MJ$
  may be much wider than $\ME$ and even contain more horizons; however, it should be kept
  in mind that conformal continuations emerge only at special values of integration constants.
  
  Simple horizons in $\ME$ with a nontrivial scalar field can certainly be obtained if the relevant 
  no-hair theorems do not work. For example, an important no-hair theorem \cite{ad-pier, herd15} 
  claims that (in GR) the domain of outer communication of an \asflat\ \bh\ cannot contain a 
  non-constant canonical minimally coupled scalar field $\psi$ with a potential $V(\psi) \geq 0$. 
  However, examples of such \bhs\ where $V\!< \! 0$ in at least a part of the range of $\psi$ do exist, 
  see, e.g., \cite{shik02}. Meanwhile, for HMPG, relevant are results valid for phantom scalars, for 
  which there is no such restriction, and \bhs\ can in principle appear even with $V \geq 0$; it is 
  known that phantom fields can make \bhs\ globally regular \cite{pha1, pha2, cle09}. 
  Numericalally obtained examples of \bhs\ in HMPG with a Higgs-like potential have been 
  considered in \cite{dan19}, and further work in this direction should also be of interest. 

\subsection{The stability problem}
  The stability properties of static configurations in STT under small perturbations may be 
  studied, as well as their geometries themselves, with the aid of field equations transformed 
  to the Einstein frame, in which there are quite numerous results concerning the fate of
  perturbations of various scalar-vacuum space-times, see, e.g., 
  \cite{b-kh79,hay02,ggs1,ggs2,kb-stab11,kb-stab12,kor15,kor17,chap17}. 
  It is clear that perturbations in Jordan's frame are governed by the same equations 
  as in Einstein's, though being expressed in other variables after the substitution \rf{J-E}. 
  What can significantly change after this transformation are the boundaries and the boundary 
  conditions that should now be imposed by physical requirements formulated in $\MJ$. 
  
  Let us briefly outline the expected results on the stability properties of the solutions 
  considered here with respect to radial perturbations (a more thorough analysis is postponed 
  for the future). 
  
  To begin with, in the case of \sph\ (monopole) perturbations of static scalar-vacuum space-times, 
  the only dynamic degree of freedom is related to scalar field perturbations $\delta\phi(u,t)$, or those 
  of the Einstein-frame field, $\delta\psi(u,t)$, because the tensor degrees of freedom only begin 
  with the quadrupole. Accordingly, the perturbations are governed by a single linear equation 
  for $\delta\psi$ while all the accompanying  perturbations $\delta\alpha, \delta\beta,
  \delta\gamma$ of the metric functions (in the notations of the metric \rf{ds_E})  
  can be excluded from this ``master equation'' with the help of the Einstein equations. 
  The resulting single equation can be written for a spectral component 
  $\delta\psi = \Psi(u) \e^{i\omega t}$ as the Schr\uml{o}dinger-like equation  
\beq                    \label{Schr}
		\frac{d^2 Y}{dz^2}  + \big(\omega^2 - W(z)\big) Y =0.		
\eeq    
  Here, $z$ is the so-called tortoise coordinate related to an arbitrary coordinate $u$ (in the 
  notations of \rf{ds_E}) by $du/dz = \e^{\gamma-\alpha}$, the unknown function is
  $Y(z) = \Psi(u) e^\beta$, and the effective potential $W(z)$ is given by
  \cite{kb-stab11,kor15,kor17}
\bearr
		 W(z) = \e^{2\gamma} \bigg[\frac{3\psi'^2}{\beta'^2}(2\e^{2\beta}\! - \! U)
			   + \frac{\psi'}{\beta'} U_\psi - \frac 1{12} U_{\psi\psi}	  \bigg]  
\nnn		\cm  
		  + \e^{2\gamma-2\alpha}[\beta'' + \beta'(\beta'+\gamma'-\alpha')],
\ear 
  where the index $\psi$ stands for $d/d\psi$, the prime for $d/du$ ($u$ is an arbitrary 
   radial coordinate in the background metric \rf{ds_E}), and $U = U(\psi) = V(\phi)/(1+\phi)^2$, 
   $V(\phi)$ being the scalar field potential in the original theory \rf{S1}. 
   
   In the solutions under study, \eqs \rf{ds_J}--\rf{int-1}, all branches A-C contain throats ($\beta'=0$) 
   in their Einstein-frame manifolds $\ME$ (even though not all of them correspond to \whs), hence 
   the potential $W(z)$ contains a singularity due to $\beta'$ in the denominator. This singularity 
   admits regularization by  replacing $W(z)$ with a new potential $W_{\rm reg}(z)$ which is 
   finite in the whole range of $u$ (or $z$) and suitable for considering the relevant boundary 
   value problems, see detailed descriptions in \cite{ggs1,kb-stab11,kb-stab12,chap17}. 
   
\begin{figure}   
\centering
\includegraphics[width=8cm]{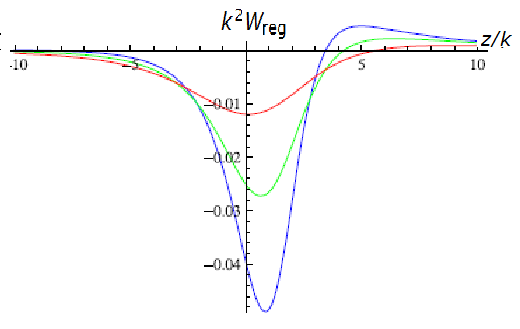}  
\caption{\small
	         Examples of $W_{\rm reg}(z)$ for branch A of the anti-Fisher solution ($k > 0$,
	         bottom-up --- $h/k = 3/2, 2, 3$)    \cite{kb-stab11}. Similar potential wells are
	         present in $W_{\rm reg}(z)$ for other branches.
	}
\end{figure}   
    
   The particular form of $W_{\rm reg}(z)$ is different for different branches of the solution 
   \rf{ds_J}--\rf{int-1} and will not be presented here, see \cite{kb-stab11}. It is important
   that all branches of the anti-Fisher solution turn out to be unstable \cite{ggs1, kb-stab11},
   the instability being connected with a potential well in $W_{\rm reg}(z)$ (see examples 
   in Fig.\,5).  Let us discuss, which of these instability conclusions can be extended to the
   present HMPG solutions in the Jordan frame, for which $W_{\rm reg}(z)$ is the same but
   the range of $z$ and the boundary conditions may be different. 
   
   For branches A and B ($k \geq 0$), in $\ME$ we have $u > 0$ corresponding to 
   $z \in \R$, and an unstable mode is found \cite{kb-stab11} under the boundary conditions 
   $\delta\psi \to 0$ as $z\to \pm \infty$.  In $\MJ$, due to the factor $\cos^2\psi$ in the metric, 
   the range of $u$ extends from zero to the singular point $u_s$ such that 
   $\psi = Cu_s + \psi_0 = \pi/2$, and the range of $z$ is also diminished: $z \in (z(u_s), \infty)$. 
   Therefore, the instability conclusion cannot be directly extended to $\MJ$, and a separate  
   study is necessary. Its results must depend on the solution parameters (including $\psi_0$)
   and on the adopted boundary condition at the singularity.
   
   The same argument applies to branch C1, also  with a singularity caused by $\cos\psi =0$.
   On the contrary, in branches C2 (a \bh) and C3 (a \wh), the whole \wh\ space-time $\ME$
   maps in $\MJ$ either to a \wh\ or to a static region of a \bh\ from the horizon to infinity. 
   Therefore, if we adopt for the horizons the same boundary condition $\delta\psi =0$ as is 
   used at infinity (which looks reasonable), then the instability result for \sph\ perturbations 
   also extends to $\MJ$. The same is true for each static region between horizons in the 
   solution described in Sec. 3.3. 
      
   This brief analysis argues that all regular solutions considered here are unstable.   
      
 \section{Concluding remarks}
      
   Considering exact vacuum solutions of HMPG with zero potential $V$, we have found that   
   many of them describe space-times with naked singularities, other generic solutions 
   correspond to traversable \whs, while only solutions of a special two-parameter family
   contain extremal Killing horizons and describe globally regular \bhs\ with zero Hawking 
   temperature. These results disagree with some of those obtained in \cite{dan19}, where 
   the same set of equations was solved numerically directly in Jordan's frame, and the 
   existence of \bhs\ with finite temperature was stated. The reason for such a discrepancy 
   is yet to be understood. An independent analysis of solutions with nonzero potentials 
   can probably shed light on this question.
      
   The \bh\ solutions obtained here are also of interest as phantom counterparts of those 
   with a usual conformal scalar \cite{kb70, bek74}. However, a tentative analysis indicates 
   that these new solutions are unstable under monopole perturbations. One may recall that 
   it was first concluded \cite{kb78} that the \bh\ solutions of \cite{kb70, bek74} are also unstable 
   but a later and more thorough analysis \cite{turok} revealed their stability. 
   
   The \wh\ configurations described here are quite a natural consequence of the phantom 
   nature of the action \rf{S1}. It is therefore not surprising that examples of \wh\ solutions with
   matter respecting the standard energy conditions were previously obtained in
   \cite{cap12, rosa18} since a violation of the Null Energy Condition, which is necessary 
   for their existence, was actually provided by the scalar $\phi$, and, as we saw, \whs\ with 
   two flat infinities are obtained even without matter. Though, precisely as their counterparts 
   in GR (anti-Fisher \whs), they turn out to be unstable since their stability 
   analysis directly extends in this case to the Jordan frame. 
   
   In the massless case ($V{=}0$) it is quite straightforward to include 
   into consideration electromagnetic fields $F\mn$ in the HMPG framework, in full analogy with 
   studies in other scalar-tensor theories \cite{kb73, kb99}. Solutions with $F\mn$ and nonzero 
   potentials can also be studied both analytically and numerically, for example, by analogy with 
   \cite{we12, kor15}.  
   
   Lastly, it is of interest to extend the present study to the so-called extended HMPG with 
   functions $f(R, \cR)$ of two curvatures \cite{boh13, mont19}, this work is in progress.
          
\subsection*{Acknowledgments}

  I thank Sergei Bolokhov, Milena Skvortsova and Vladimir Ivashchuk for helpful discussions. 
  The work was partly performed within the framework of the Center FRPP 
  supported by MEPhI Academic Excellence Project 
  (contract No. 02.a03.21.0005, 27.08.2013).
  The work was also partly funded by the RUDN University Program 5-100
  and the Russian Basic Research Foundation grant 19-02-0346.
     

\end{document}

%% file: preamble.tex
\usepackage{amsfonts,amssymb,cite}

\def\noi{\noindent}

\newcommand{\Title}[1]{\noi {{\Large\bf #1}}\\[1ex]}

\newcommand{\Author}[2]{\noi{\bf #1}\\[2ex]\noi{\normalsize\it #2}\\}

\newcommand{\Abstract}[1]{\vskip 2mm \begin{center}
        \parbox{16.4cm}{\small\noi #1} \end{center}\medskip}

\newcommand{\foom}[1]{\protect\footnotemark[#1]}

\def\email#1#2{\footnotetext[#1]{e-mail: #2}\addtocounter{footnote}{1}}

\def\nq{\hspace*{-1em}}
\def\nqq{\hspace*{-2em}}
\def\nhq{\hspace*{-0.5em}}

\def\cm{\hspace*{1cm}}
\def\inch{\hspace*{1in}}

\def\Jl#1#2{#1 {\bf #2},\ }

\def\ApJ#1 {\Jl{Astroph. J.}{#1}}
\def\CQG#1 {\Jl{Class. Quantum Grav.}{#1}}
\def\DAN#1 {\Jl{Dokl. AN SSSR}{#1}}
\def\GC#1 {\Jl{Grav. Cosmol.}{#1}}
\def\GRG#1 {\Jl{Gen. Rel. Grav.}{#1}}
\def\JETF#1 {\Jl{Zh. Eksp. Teor. Fiz.}{#1}}
\def\JETP#1 {\Jl{Sov. Phys. JETP}{#1}}
\def\JHEP#1 {\Jl{JHEP}{#1}}
\def\JMP#1 {\Jl{J. Math. Phys.}{#1}}
\def\NPB#1 {\Jl{Nucl. Phys. B}{#1}}
\def\NP#1 {\Jl{Nucl. Phys.}{#1}}
\def\PLA#1 {\Jl{Phys. Lett. A}{#1}}
\def\PLB#1 {\Jl{Phys. Lett. B}{#1}}
\def\PRD#1 {\Jl{Phys. Rev. D}{#1}}
\def\PRL#1 {\Jl{Phys. Rev. Lett.}{#1}}

\def\lal{&&\nqq {}}
\def\eq{Eq.\,}
\def\eqs{Eqs.\,}
\def\beq{\begin{equation}}
\def\eeq{\end{equation}}
\def\bear{\begin{eqnarray}}
\def\bearr{\begin{eqnarray} \lal}
\def\ear{\end{eqnarray}}
\def\earn{\nonumber \end{eqnarray}}

\def\nnn{\nonumber\\ \lal }

\def\yy{\\[5pt] {}}
\def\yyy{\\[5pt] \lal }

\def\dst{\displaystyle}
\def\tst{\textstyle}
\def\fracd#1#2{{\dst\frac{#1}{#2}}}
\def\fract#1#2{{\tst\frac{#1}{#2}}}
\def\Half{{\fracd{1}{2}}}
\def\half{{\fract{1}{2}}}

\def\e{{\,\rm e}}
\def\d{\partial}

\def\sign{\mathop{\rm sign}\nolimits}

\def\const{{\rm const}}
\def\eps{\varepsilon}

\newcommand{\vars}[1]{\left\{\begin{array}{ll}#1\end{array}\right.}

\def\uml#1{\"{#1}}

